# Phase transitions in the complex plane of physical parameters

Bo-Bo Wei, Shao-Wen Chen, Hoi-Chun Po & Ren-Bao Liu*

Department of Physics, Centre for Quantum Coherence, and Institute of Theoretical Physics, The Chinese University of Hong Kong, Shatin, New Territories, China

*rbliu@phy.cuhk.edu.hk

**Abstract**

**At low temperature a thermodynamic system undergoes a phase transition when a physical parameter passes through a singularity point of the free energy, corresponding to formation of a new order. At high temperature the thermal fluctuations destroy the orders; correspondingly the free energy is a smooth function of the physical parameter and the singularity would occur only at complex values of the parameter. Since a complex parameter is unphysical, at high temperature no phase transition is expected with varying the physical parameter. Here we show that the quantum evolution, driven by a designed interaction, of a system initialized in thermal equilibrium is equivalent to the partition function of a complex parameter. Therefore we can access the complex singularity points of thermodynamic functions and observe phase transitions even at high temperatures. This discovery makes it possible to study thermodynamics in the complex plane of physical parameters.**

**Main text**

The physical properties of a many-body system at thermodynamic equilibrium are fully determined by the partition function $Z(\beta, \lambda_1, \lambda_2, \ldots, \lambda_K)$ as a function of coupling parameters $\{\lambda_k\}$ and the temperature $T$ (or the inverse temperature $\beta \equiv 1/T$). The partition function is the summation of the Boltzmann factor $e^{-\beta H(\lambda_1, \lambda_2, \ldots, \lambda_K)}$ over all energy eigen states, i.e., $Z = \text{Tr}\left[e^{-\beta H(\lambda_1, \lambda_2, \ldots, \lambda_K)}\right]$, where the Hamiltonian $H = \sum_k \lambda_k H_k$ is characterized by a set of coupling parameters $\{\lambda_k\}$ (e.g., in spin models magnetic field $h = \lambda_1$, nearest neighbor coupling $J = \lambda_2$, and next nearest neighbor coupling $J' = \lambda_3$, etc.). The normalized Boltzmann factor $Z^{-1}\exp(-\beta H)$ is the probability of the system in a state with energy $H$.

At a sufficiently low temperature ($\beta > |\lambda_k|$), the many-body system may have different orders (phases) in different parameter ranges and therefore phase transitions may occur with varying a coupling parameter across the boundaries (see Figure 1**a)**. The phase transitions correspond to non-analytic or singularity points of the free energy $F$ (which is related to the partition function by $F = -\beta^{-1}\ln Z$). For a finite-size system the partition function is always positive and hence the free energy always analytic for real physical parameters. The partition function can be zero for complex temperature or complex coupling parameters (see Figures 1**b** & 1**c)**, such as Lee-Yang zeros [1,2] in the complex plane of magnetic field for a spin lattice and Fisher zeros [3] in the complex plane of temperature. At a sufficiently low temperature and in the thermodynamic limit (the number of particles in the system approaching to infinity), the zeros of the partition function can approach to real axes of the coupling parameters where phase transitions occur [1-2] (see Figure 1**c**). At high temperature the free energy would be non-analytic only for complex values of temperature and coupling parameters (see Figure 1**b**) [3-5]. Therefore no phase transition would occur with varying a coupling parameter at high temperature (see Figure 1**a**). Physically, this is because at high temperature the thermal fluctuation destroys all possible orders and prevents the phase transitions. Phase transitions at high temperature would be possible if we could have complex parameters, which, however, are generally regarded as unphysical.

Here we present a systematic approach to complexification of an arbitrary physical parameter, using quantum evolution that involves a complex phase factor. There are two fundamental laws in physics that involve probabilistic distributions. One is the Boltzmann distribution $Z^{-1}\exp(-\beta H)$, the *probability* of a system in a state with energy $H$ at inverse temperature $\beta$. The other is the Schrödinger equation that gives $\exp(-itH)$, the *probability amplitude* of a system in a state with energy $H$ at time *t*. The unit imaginary number $i \equiv \sqrt{-1}$ is associated with the time in quantum evolution. Thus a time-dependent measurement of the system

$$L(t) = Z^{-1}\text{Tr}\left[\exp(-\beta H - itH_I)\right] \qquad (1)$$

has the form of a partition function with an imaginary coupling parameter $it/\beta$. By choosing $H_I$ to be the *k*-th component of $H = \sum_k \lambda_k H_k$, i.e., $H_I = H_k$, the *k*-th coupling parameter is continuated to the complex plane via $\lambda_k \to \lambda_k + it/\beta$. One can also realize complexification of the inverse temperature ($\beta \to \beta + it$) by choosing $H_I = H$. Such time-domain measurement, as will be discussed later in this paper, is experimentally realizable though non-trivial in general cases. The time-domain measurement in equation (1) is fully determined by the *equilibrium-state* partition function of the system. The measurement result will be zero when a zero of the partition function in the complex plane is encountered. And critical times may exist corresponding to the singularity points in the complex plane, as indicated by a recent study of central spin decoherence caused by an Ising bath [6]. Now we have a systematic approach to accessing different types of zeros in the complex plane for the partition function, such as Lee-Yang zeros by choosing $H_I = \sum_j S_j^z$ for a system of spins $\{\mathbf{S}_j\}$ [1-2], Fisher zeros by choosing $H_I = H$ [3], and other types of zeros (yet unnamed) by choosing $H_I = \sum S_j^x S_{j+1}^x$, $H_I = \sum \mathbf{S}_j \cdot \mathbf{S}_{j+1}$, etc. More importantly, by complexification of a physical parameter, we have a way to access the singularity points in the complex plane and to reveal phase transitions at high temperature ($\beta^{-1} \gg \{|\lambda_k|\}$) via time-domain measurement. The high-temperature phase transitions are also inferred from a recent finding that a quantum criticality can emerge at high temperatures by long-time quantum evolution [7].

**High-temperature magnetic phase transitions**

Spin systems can have ferromagnetic (FM) or antiferromagnetic (AFM) orders at low temperatures, corresponding to positive or negative coupling (*J*) between the spins, respectively. Thus an FM-AFM transition would occur if the coupling *J* varies from positive to negative. At high temperatures, the thermal fluctuation destroys the magnetic orders and hence no phase transition is expected with changing $J$ in a certain range. Here we study the Ising spin models to demonstrate the FM-AFM transition in the complex plane of parameters. The Hamiltonian of a general Ising model is

$$H = -\sum_{i,j} J_{ij}\sigma_i\sigma_j - h\sum_j \sigma_j, \qquad (2)$$

where $J_{ij}$ is the coupling between spins $\sigma_i$ and $\sigma_j$, $h$ is the magnetic field, and the spins $\sigma_j$ take values $\pm 1$. At low temperature, when the coupling changes from positive to negative, the Ising model presents a phase transition from FM order to AFM order at zero field. Correspondingly, the Lee-Yang zeros in the complex plane of scaled magnetic field $z \equiv \exp(2\beta h)$ exhibits different distributions. Note that the distinct features of Lee-Yang zero distributions in the complex plane persist even at high temperatures ($T >> |J_{ij}|$).

To be specific, we study the one-dimensional (1D) Ising model with nearest-neighbor coupling $J$, which can be exactly solved through the transfer matrix method [8-10] (see Supplementary Information). There is no finite temperature phase transition in the 1D Ising model. The Lee-Yang zeros of the 1D Ising model of $N$ spins have been exactly calculated [2]. We plot the distribution of Lee-Yang zeros in Figure 2**a**. For AFM coupling ($J < 0$), all the zeros lie on the negative real axis (indicated by the red surface) (Figure 2**a**). While for the FM coupling ($J > 0$), the zeros are distributed on an arc of the unit circle (indicated by the blue surface) (Fig. 2**a**). At the transition point ($J = 0$), all the Lee-Yang zeros are degenerate at $z_n = -1$ (indicated by the green solid ball).

To observe these Lee-Yang zeros and the critical behaviors in the complex plane [4-5], we study the time-domain observation in equation (1) by choosing $H_I = \sum_j \sigma_j$ to continuate the magnetic field to the complex plane $h \to h + it/\beta$. In this case, the time-domain measurement corresponds to decoherence of a quantum probe spin coupled to the Ising model [6] (see physical realizations). Figures 2**b-e** plot the time-domain measurement for different coupling parameters $J$ at fixed inverse temperature $\beta = 1$ (which is a high temperature case for the 1D Ising model, since the critical temperature of this model is zero). For AFM coupling $J = -1$, the time-domain measurement has no zeros (Figure 2**b**) and it is a smooth function of time. On the contrary for the FM coupling $J = 1$, the time-domain measurement shows a number of zeros (Figure 2**c**), which one-to-one correspond to the Lee-Yang zeros [6]. Approaching to the thermodynamic limit, the Yang-Lee edge singularities (the starting and ending Lee-Yang zeros along the arcs) [4-5] lead to critical times in the time-domain observation [6]. To demonstrate this, we do finite size scaling analysis on the time-domain measurement and show the scaled results $|L(t)|^{1/N}$ in Figures 2**d** & 2**e**. The profiles of the time-domain measurement in the FM and AFM regions are qualitatively different. For AFM coupling, the scaled measurement is a smooth function of time (Figure 2**d**). While for FM coupling, the scaled measurement presents sudden changes at critical times corresponding to the Yang-Lee edge singularities (Figure 2**e**). The profiles of the time-domain measurement in the FM and AFM regions cannot be smoothly transformed into each other, which signatures the onset of a high-temperature phase transition.

We further study the high-temperature AFM-FM phase transition in time-domain measurement for a two-dimensional (2D) Ising model. Specifically, we consider a 2D Ising model in a square lattice with nearest neighbor coupling $J$. This model under zero field is exactly solvable [9,10] and has a finite-temperature phase transition at $\beta_C \approx 0.44/|J|$. Figure 3 shows the time-domain measurement in the 2D Ising model for different coupling parameters $J$ at a fixed high temperature, $\beta = 0.3 < \beta_C$. For the AFM coupling $J = -1$, the partition function has no Lee-Yang zeros on the unit circle ($|z| \equiv |\exp(2\beta h)| = 1$)[11] and therefore the time-domain measurement has no zeros (Figure 3**a**). While for the FM coupling $J = 1$, the time-domain measurement presents a number of zeros (equal to the number of spins) (Figure 3**b**), corresponding to the Lee-Yang zeros along the unit circle. We do finite-size scaling analysis in Figure 3**c** and Figure 3**d** for AFM coupling and FM coupling, respectively. It is clear that the time-domain measurement is a smooth function of time for the AFM coupling, while it presents sudden changes at critical times corresponding to the Yang-Lee edge singularities for the FM coupling (Figure 3**d**). Thus a phase transition with varying the coupling constant $J$ occurs at a temperature higher than the critical temperature ($\beta < 0.44/|J|$).

**Renormalization group theoretic analysis**

The renormalization group (RG) theoretic methods, a powerful tool for studying conventional phase transitions, can be applied to the phase transitions in the complex plane of physical parameters. Since the phase of any complex number is defined modulo $2\pi$, the RG flows of complex parameters can present novel topological structures.

As an example we first consider the 1D Ising model and define the dimensionless parameters $K_0 = \beta J$ and $h_0 = \beta h$. The renormalization of the model can be exactly formulated by blocking two neighboring spins into one (Figure 4**a**) [12]. By continuation of the dimensionless external field to a purely imaginary value $h_0 = i\tau_0 / 2$, the exact RG flow equations [12] become

$$\begin{cases} \tau_1(\tau_0, K_0) = \tau_0 - i\ln\left(\dfrac{\cosh(2K_0 + i\tau_0/2)}{\cosh(2K_0 - i\tau_0/2)}\right) \\ K_1(\tau_0, K_0) = \dfrac{1}{4}\ln\left(\dfrac{\cosh(4K_0) + \cos\tau_0}{1 + \cos\tau_0}\right) \end{cases}, \qquad (3)$$

where the coupling $K_1$ remains real after renormalization. Physically, this specification of parameters corresponds to the time domain measurement of the system in zero external field. Since $\tau$ is defined modulo $2\pi$, the parameter space can be identified with the surface of an infinitely long cylinder with unit radius. The original system corresponds to the curve $K = K_0$ and $-\pi < \tau_0 \leq \pi$, and therefore its winding number ($W_\#$), defined to be the number of times a closed curve wrapped on the cylinder, is 1 (Fig. 4**b**).

The original parameter curve is renormalized according to the RG flow equations. Since $K_1(\tau_0, K_0) = K_1(\tau_0, -K_0)$ in Eq. (3), the distinct behaviors of the FM and AFM

Ising chains are encoded entirely in different RG flows of $\tau$. This is illustrated in Fig. 4**c-h**. Figures 4**c**, 4**e** & 4**g** show that after successive renormalization, the winding number in the FM case ($K_0 = 1/8$) becomes 2, 4, 8 .... On the contrary, Figures 4**d**, 4**f** & 4**h** show that after successive renormalization, the winding number in the AFM case ($K_0 = -1/8$) is zero. The RG flow equations become trivial at the phase transition point $K_0 = 0$, which corresponds to the infinite temperature limit, and the winding number remains unchanged ($W'_\# = 1$) after the renormalization. In summary, the winding numbers for different couplings after $k$ steps of renormalization are

$$W_\# \to W_\#^{(k)} = \begin{cases} 0 & \text{for} \quad K_0 < 0 \\ 1 & \text{for} \quad K_0 = 0. \\ 2^k & \text{for} \quad K_0 > 0 \end{cases} \qquad (4)$$

The different topologies of the RG flows demonstrate unambiguously the high-temperature phase transition with varying the coupling parameter.

We further consider the 2D Ising model in a square lattice. By continuation of the external field to a purely imaginary value of $h = i\tau/2$, the approximate RG flow equations [13] read (See Supplementary Information for derivation)

$$\begin{cases} \tau_1(\tau_0, K_0) = \tau_0 - \dfrac{i}{2} \ln\left( \dfrac{\cosh(4K_0 + i\tau_0/2)\cosh^2(2K_0 + i\tau_0/2)}{\cosh(4K_0 - i\tau_0/2)\cosh^2(2K_0 - i\tau_0/2)} \right) \\ K_1(\tau_0, K_0) = \dfrac{3}{16} \ln\left( \dfrac{\cosh(8K_0) + \cos(\tau_0)}{1 + \cos(\tau_0)} \right) \end{cases}. \qquad (5)$$

Figure 5 presents the RG flows of the parameters in FM and AFM cases. Figures 5**a**, **c** & **e** present the renormalized parameters under renormalization once, twice and three times in turn for the FM case ($K_0 = 1/8$). Figures 5**b**, **d** & **f** present the results for the

AFM case ($K_0 = -1/8$). The winding numbers of the different cases after $k$ steps of renormalization are

$$W_{\#} \to \begin{cases} \left[(5/2)^k\right] & \text{for } K_0 > 0 \\ 1 & \text{for } K_0 = 0, \\ \left[(5/2)^{k-1}/2\right] & \text{for } K_0 < 0 \end{cases} \quad (6)$$

where $[x]$ is the integer part of $x$. The winding number of the parameters under RG reflects the different topology intrinsic to the RG flow equations in the different parameter regimes.

**Transverse-field Ising model**

The models considered above are all classical models in which different components of the Hamiltonian commute. A natural question arises about whether the high-temperature phase transitions with complex parameters would exist also for quantum models. To address this question, we study the 1D transverse-field Ising model. The model contains $N$ spin-1/2 with nearest neighbor interaction $(\lambda_1)$ along the $x$-axis and under a transverse field $(\lambda_2)$ along the $z$-axis, described by the Hamiltonian

$$H = \lambda_1 H_1 + \lambda_2 H_2, \, H_1 = \sum_{j=1}^{N} \sigma_j^x \sigma_{j+1}^x, \, H_2 = \sum_{j=1}^{N} \sigma_j^z, \quad (7)$$

where $\sigma_j^{x/y/z}$ is the Pauli matrix of the $j$-th spin along the $x/y/z$-axis. This model is exactly solvable [14]. It has a quantum phase transition between a magnetic ordered phase for $|\lambda_1| > |\lambda_2|$ and a disordered phase for $|\lambda_1| < |\lambda_2|$ at zero temperature, but has no finite-temperature phase transition for any parameters on the real axis. By defining the dimensionless magnetic field $h = \lambda_2/\lambda_1$, the Lee-Yang zeros are determined by $\text{Re}(h)^2 + \text{Im}(h)^2 = 1 + [(n+1/2)\pi/\beta]^2$, $|\text{Re}(h)| \leq 1$. Therefore, the zeros are located on circles and have cutoff at singularity edges $\text{Re}(h) = \pm 1, \text{Im}(h) = \pm|(n+1/2)\pi/\beta|$ (see Figure 6a). When the temperature approaches zero $(|\beta| \to \infty)$, the Lee-Yang zeros are on the unit circle. When the temperature is high $(|\beta| \ll 1)$, the radii of the circles $\sqrt{1+|(n+1/2)\pi/\beta|^2} \gg 1$. Therefore, the zeros are distributed, approximately, on horizontal lines with interval $|\pi/\beta|$. The fact that the Lee-Yang zeros exist only in the parameter range of $|h| \leq 1$ indicates that the time-domain measurement of the system would present phase transitions between the two parameter regions, $|\lambda_2| > |\lambda_1|$ and $|\lambda_2| < |\lambda_1|$. Specifically, the time-dependent measurement can be devised as $L(t) = Z^{-1}\text{Tr}[\exp(-\beta H - itH_2)]$, which is the partition function with a complex external field ($\lambda_2 \to \lambda_2 + it/\beta$). The contour plot of the time-domain measurement as a function of external field and time are presented in Figure 6b. To demonstrate the phase transitions more clearly, we plot in Figure 6c the time-domain measurement as a function of external field for different times. Since the zeros are bounded in the range $|h| \leq 1$ with Yang-Lee edge singularities at $\text{Re}(h) = \pm 1, \text{Im}(h) = \pm|(n+1/2)\pi/\beta|$, the time-domain measurement has a sharp change when we tune the parameter from $|h| \leq 1$ to $|h| > 1$ at times $\text{Im}(h) = \pm|(n+1/2)\pi/\beta|$ (See Figure 6c).

**Physical realization**

The time-domain measurement in equation (1) resembles the Loschmidt echo [15,16], or equivalently, decoherence of a central spin coupled to the system. Thus we may implement the time-domain measurement by coupling the system to a central spin through the probe-system coupling $|\uparrow\rangle\langle\uparrow|\otimes H_\uparrow(t)+|\downarrow\rangle\langle\downarrow|\otimes H_\downarrow(t)$ ($S_z \equiv |\uparrow\rangle\langle\uparrow|-|\downarrow\rangle\langle\downarrow|$) and measuring the central spin coherence. Essentially, the coherence of the probe spin is a complex phase factor associated with a real Boltzmann probability for each state of the system. Therefore the probe spin coherence measurement amounts to continuation of a physical parameter to the complex plane. If we initialize the central spin in a superposition state $|\uparrow\rangle+|\downarrow\rangle$ and the system in a thermal equilibrium state described by the canonical density matrix, $\rho = Z^{-1}\text{Tr}[\exp(-\beta H)]$, the probe spin coherence is

$$\langle S_+\rangle = \langle S_x\rangle + i\langle S_y\rangle = \text{Tr}\left[e^{-\beta H}e^{i(H+H_\uparrow)t}e^{-i(H+H_\downarrow)t}\right]\Big/\text{Tr}\left[e^{-\beta H}\right], \tag{8}$$

If $[H_I, H] = 0$, the time-domain measurement $L(t) = Z^{-1}\text{Tr}\left[\exp(-\beta H)\exp(-iH_I t)\right]$, which equals to the probe spin coherence with a probe-bath coupling $H_{SB} = S_Z \otimes H_I$, i.e., $H_\uparrow = -H_\downarrow = H_I/2$. Or if $[H_I, H] \neq 0$ but $[[H_I, H], H] = [[H_I, H], H_I] = 0$, the time-domain measurement can also be factored as $L(t) = Z^{-1}\text{Tr}\left[\exp(-\beta H)\exp(-iH_I t)\right]$ (See Supplementary Information for details) and can be implemented by the probe spin coherence with a modified probe-bath coupling $H_\downarrow = H_I/2 - H$ & $H_\uparrow = -H_I/2 - H$. If $[H, H_I] = isH_I$, the time-domain measurement can be written as

$L(t) = Z^{-1}\text{Tr}[\exp(-it\sin(s)H_I/s)\exp(-\beta H)]$ (See Supplementary Information for

details), which can be implemented by probe spin decoherence with a probe-bath coupling $H_\downarrow = \sin(s)H_I/2s - H$ & $H_\uparrow = -\sin(s)H_I/2s - H$. In general cases, the time-domain measurement in equation (1) can be written as

$$L(t) = Z^{-1}\text{Tr}\left[e^{-\beta H'}e^{-itH'_I}\right], \text{ with}$$

$$e^{-2\beta H'} \equiv T\exp\left(-\frac{\beta}{2}\int_{-1}^{1}\tilde{H}(u)du\right)\overline{T}\exp\left(-\frac{\beta}{2}\int_{-1}^{1}\tilde{H}(u)du\right),$$

$$\exp(-itH'_I) \equiv \exp(\beta H')\exp(-\beta H - itH_I) \text{ and}$$

$\tilde{H}(u) \equiv \exp(iutH_I/2)H\exp(-iutH_I/2)$, where $T$ and $\overline{T}$ are the time-ordering and anti-ordering operators, respectively. If one initializes the bath in a canonical state $\rho = \exp(-\beta H')/\text{Tr}[\exp(-\beta H')]$, the time-domain measurement can be implemented by probe spin decoherence with probe-bath coupling $H_\downarrow = H'_I/2 - H'$ & $H_\uparrow = -H'_I/2 - H'$, up to a normalization factor (See Supplementary Information for details). The physical realization of the modified Hamiltonians $H'$ and $H'_I$ is non-trivial.

Note that the time-domain measurement in equation(1) is similar to the measurement of the characteristic function of the work distributions in a quantum quench [17-19], which plays a central role in the fluctuation relations in non-equilibrium thermodynamics [20]. The probe decoherence realization of the time-domain measurement can also be related to the quench dynamics where the evolution of the system can be controlled under different Hamiltonians for different periods of time [21].

**Summary**

We have shown that the quantum evolution of a system triggered from a thermodynamic equilibrium is equivalent to the partition function of the system with a complex parameter. By choosing different coupling forms we have a systematic way to realize continuation of an arbitrary physical parameter to the complex plane. The time-domain measurement allows us to study all kinds of zeros of the partition function. More importantly, we can access the singularity points of thermodynamic functions in the complex plane of physical parameters and therefore observe phase transitions at high temperatures. The physical realization of the time-domain measurement may be nontrivial but in principle it may be implemented by probe spin decoherence or quantum quench experiments. This discovery makes it possible to study thermodynamics in the complex plane of physical parameters.

## Methods

The 1D Ising spin model was exactly diagonalized by the transfer matrix method [8,9]. The evaluation of the partition function after the transformation becomes a trivial problem of diagonalization of a 2x2 matrix. The probe spin coherence was similarly calculated (which had been formulated in terms of partition functions [6]). Similarly, by the transfer matrix method, the 2D Ising model without magnetic field was mapped to a 1D Ising model with a transverse field [8,9]. For a 2D Ising model in finite magnetic field, it was mapped by transfer matrix method to a 1D Ising model with both longitudinal field and transverse field [9], which was numerically diagonalized. Therefore the partition function and hence the probe spin coherence for the 2D Ising model in finite magnetic field were obtained. We derived the exact RG equation of 1D Ising model and approximate RG equations for 2D Ising model in square lattice for real parameters [12,13]. By analytic continuation we obtained the RG equations in the complex plane of the physical parameters and analyzed the RG flow of the complex parameters. The 1D Ising model with a transverse magnetic field was exactly solved [14] and the partition function and time-domain measurement were then calculated.

Full methods and related references are included in the Supplementary Information.

**Acknowledgements:** This work was supported by The Chinese University of Hong Kong Focused Investments Scheme, and Hong Kong Research Grants Council - Collaborative Research Fund Project HKU10/CRF/08.

**Author Contributions**: R.B.L. conceived the idea, designed project, formulated the theoretical formalism, and supervised the project. B.B.W. studied magnetic phase transitions in the Ising models and calculated the RG flows of the 2D Ising model. H.C.P. discovered the topological features of the RG flows of the 1D model. S.W. studied the transverse-field Ising model. B.B.W. & R.B.L. wrote the manuscript. All authors discussed the results and the manuscript.

**Competing financial interests** The authors declare no competing financial interests.

**Correspondence** and requests for materials should be addressed to R.B.L. (rbliu@phy.cuhk.edu.hk).

**Supplementary Information** is linked to the paper.

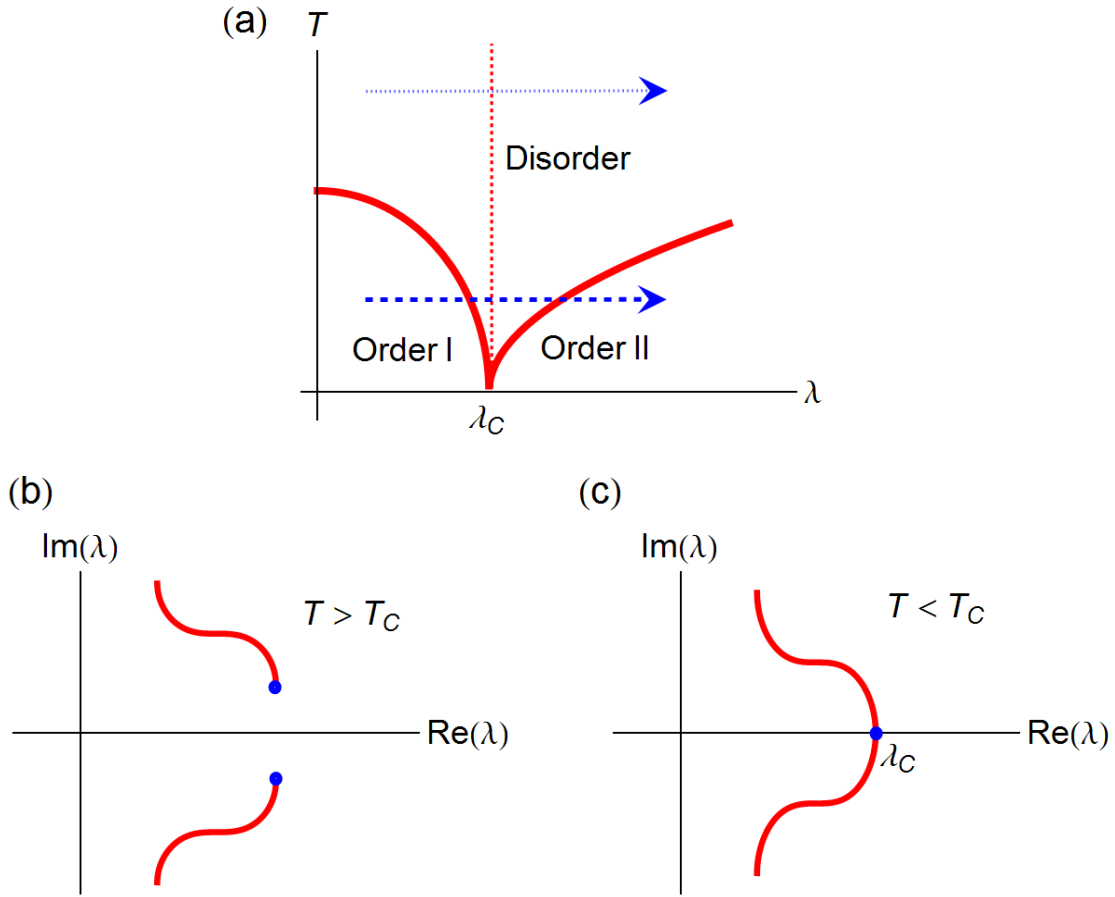

**Figure 1 |Phase transitions and zeros of a partition function in the complex plane of a physical parameter**. **a**. The schematic phase diagram of a general many-body system. At low temperature, the system has two ordered phases, I and II, separated by a critical parameter $\lambda_C$. At high temperature, the system is in the disordered phase. When we tune the control parameters at low temperatures (as indicated by blue dashed arrow) we cross two phase boundaries (indicated by solid-red lines) and therefore experience two phase transitions. Conversely at high temperature sweeping the parameter in the same range (blue dotted arrow) would not cause a phase transition. **b**. The schematic distribution of zeros of the partition function in the complex plane of the parameter for temperature $T$ above the critical point $T_C$. The two blue points mark the edge singularities. **c**. The same as **b** but for the temperature

below the critical point ($T < T_C$). The singularity edges approach to the real axis in the thermodynamic limit.

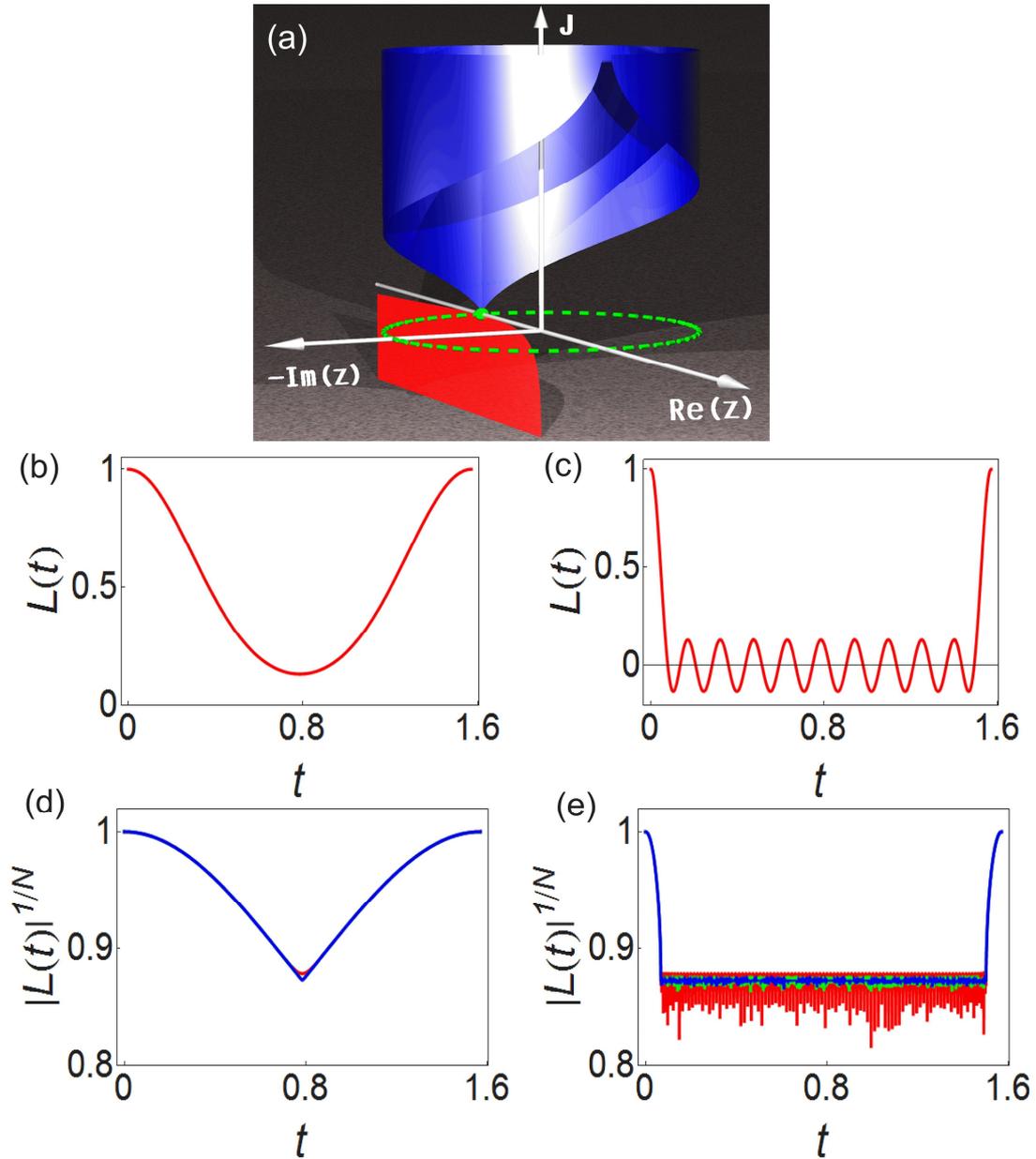

**Figure 2 | High-temperature AFM-FM phase transitions revealed by time-domain observation. a.** Schematic 3D plot of the Lee-Yang zeros in the 1D Ising model. For $J>0$, the Lee-Yang zeros are distributed on an arc of the unit circle with two edge singularity points determined by the coupling strength (blue surface). With coupling strength decreases the arcs shrink and finally the singularity edges merge to a point at $z=-1$ for $J=0$ (indicated by the green solid ball). For $J<0$, the Lee-Yang zeros lie on the negative real axis (red surface). **b.** The time-domain

measurement as a function of time for 1D Ising model with $N = 20$ spins at $\beta = 1$ and $J = -1$ (AFM). **c**. The same as **b** but for $J = 1$ (FM). **d**. Finite size scaling of the time-domain measurement in the 1D Ising model at $\beta = 1$ and $J = -1$ (AFM), the red line is for $N = 100$ spins, green line for $N = 500$ and blue line for $N = 1000$; **e**. The same as **d** but for $J = 1$ (FM).

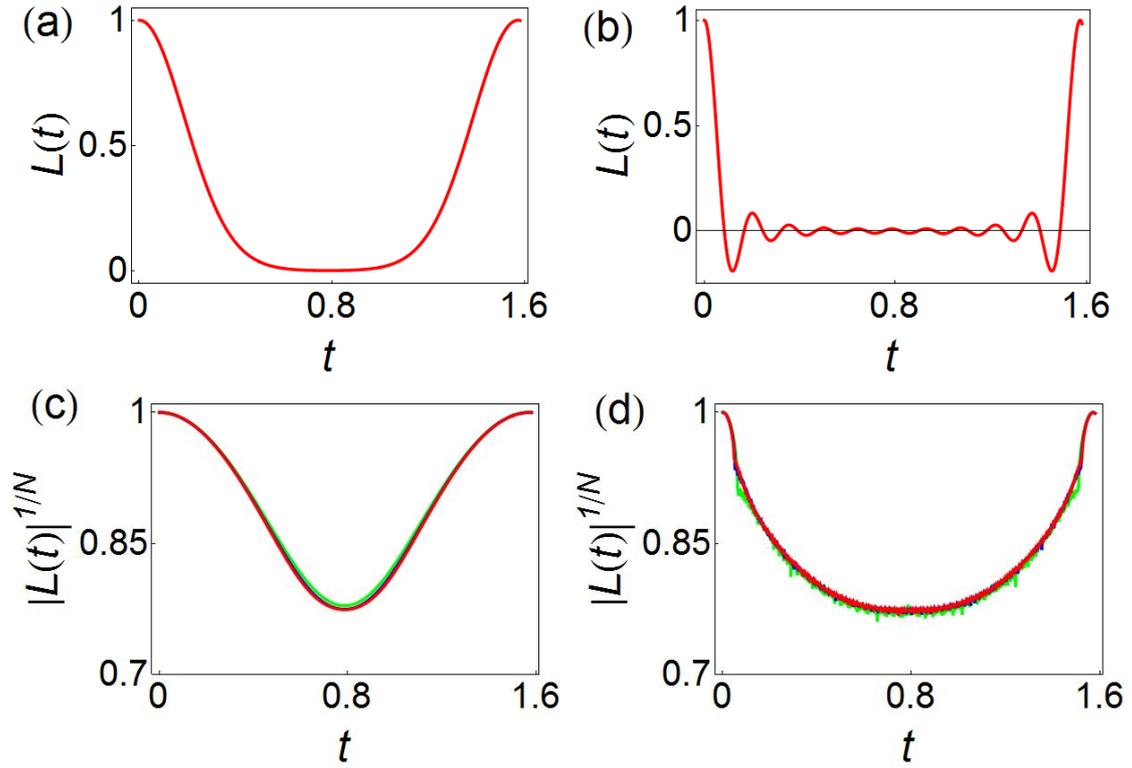

**Figure 3 | High-temperature AFM-FM phase transitions of the 2D Ising spin model in a square lattice revealed by time-domain observation. a.** The time-domain measurement as a function of time for the 2D Ising model in a $N = 4 \times 5$ square lattice, $\beta = 0.3$ and $J = -1$ (AFM). **b**. The same as **a** but for $J = 1$ (FM). **c**. Finite size scaling of the time-domain measurement with $\beta = 0.3$ and $J = -1$ (AFM), the green line is for $N = 4 \times 100$ spins, the blue line is for $N = 6 \times 100$ and red line for $N = 8 \times 100$. **d**. The same as **c** but for $J = 1$ (FM).

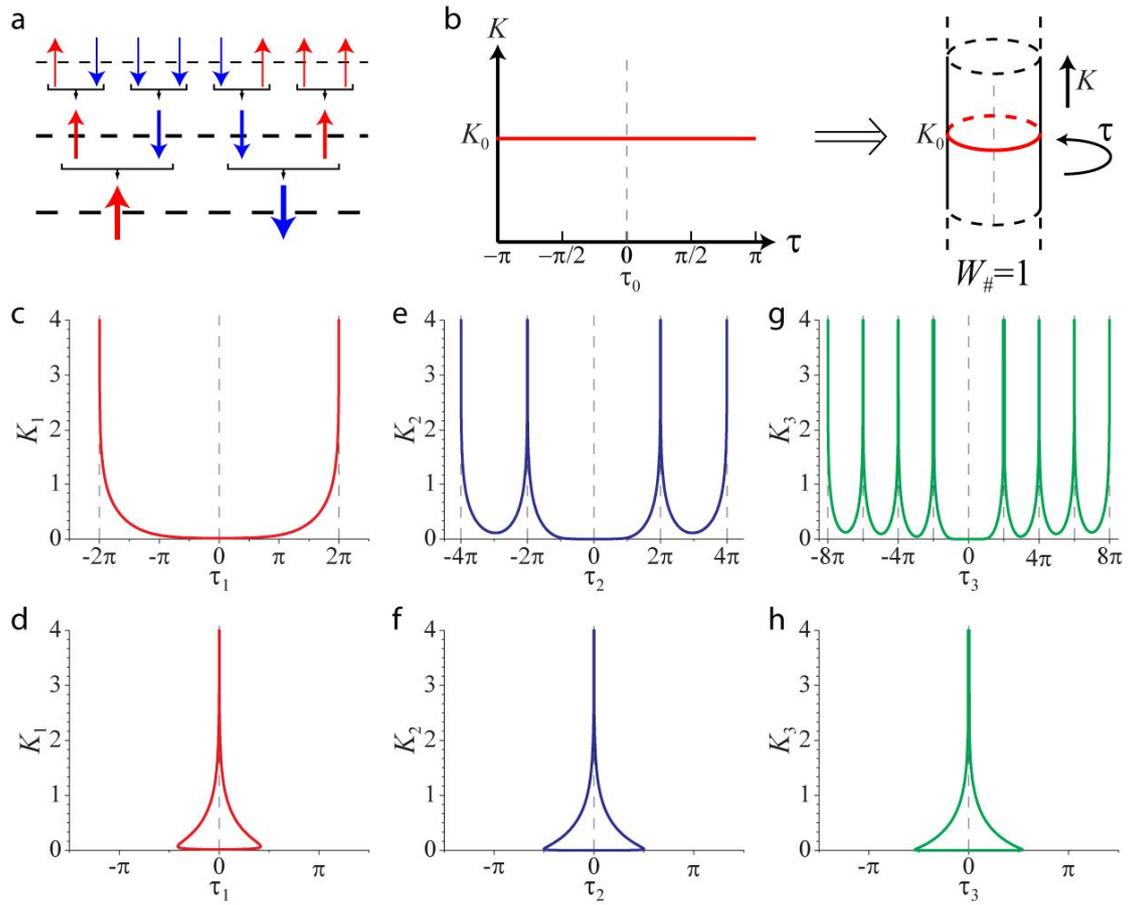

**Figure 4 | Renormalized parameters of the 1D Ising model with a purely imaginary field. a,** Real-space renormalization scheme. In each renormalization step the spins on every other site are traced out, which effectively combines pairs of neighboring spins into block spins with renormalized coupling and external field. **b,** The un-renormalized parameters. The spin coupling $K$ and evolution time $\tau$ (i.e., imaginary part of the external field) form a space identical to the surface of an infinitely long cylinder. When the imaginary field $\tau$ is varied at a fixed value of $K = K_0$, the curve winds about the cylinder once and so the winding number $W_\# = 1$. **c-h,** Under RG flow, the original curve depicted in (**a**) would transform differently in the different parameter regimes of $K_0$. (**c**), (**e**) and (**g**) in turn show the parameters for the FM case with $K_0 = 1/8$ after one, two, and three steps of renormalization. (**d**),

(**f**) and (**h**) show the corresponding results for the AFM case with $K_0 = -1/8$. The vertical dashed lines indicate $\tau = 0 \mod(2\pi)$, which are identified as the same line when represented on an infinitely long cylinder. The winding numbers can be directly inferred from the number of times the renormalized curve crosses this line.

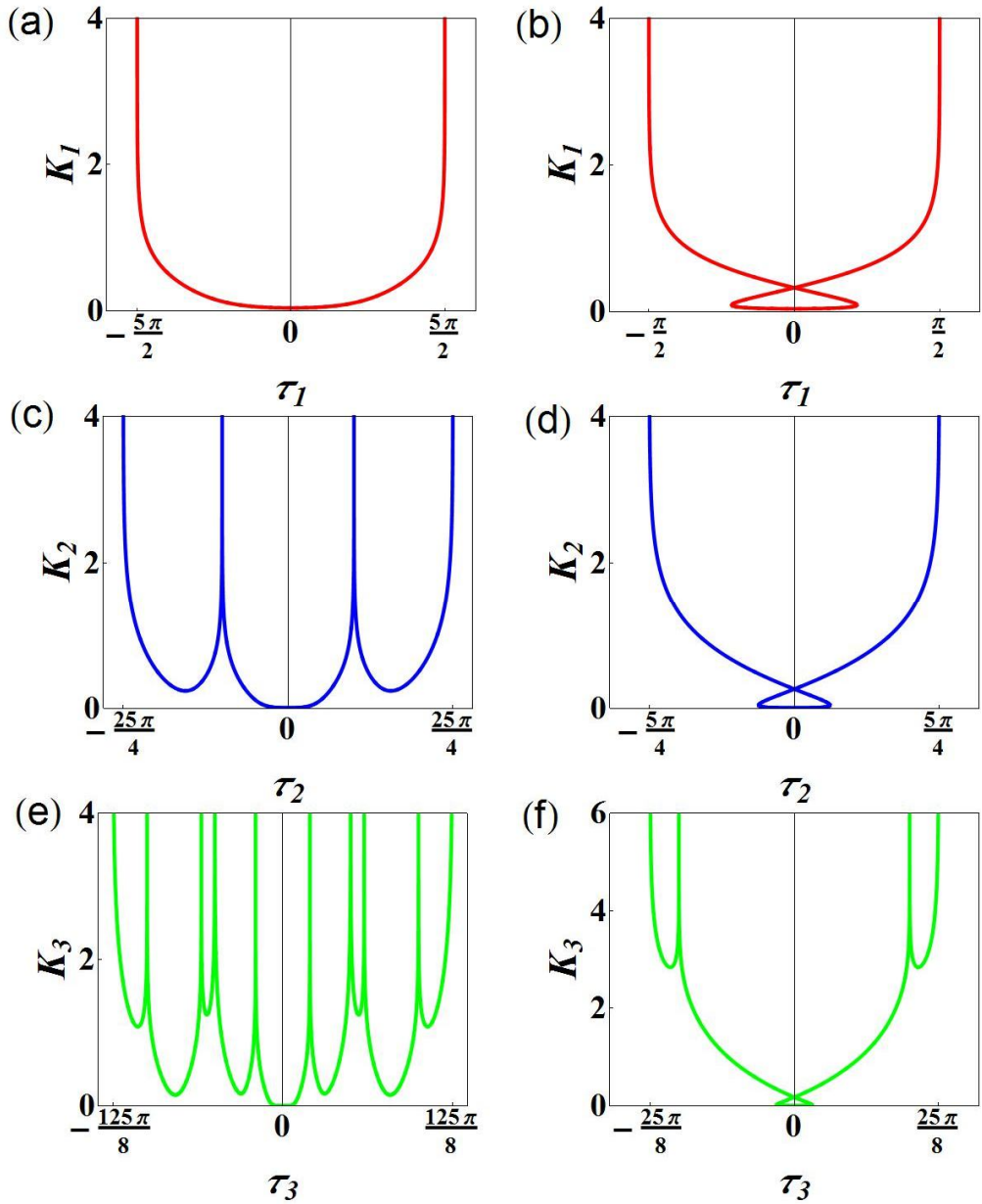

**Figure 5 | Topological difference between the renormalization group flows of the FM and AFM 2D Ising models.** (**a**), (**c**) & (**e**) present in turn the parameters (spin coupling and imaginary external field) after once, twice and three times of renormalization in the FM case ($K_0 = 1/8$). (**b**), (**d**) & (**f**) are the same as (**a**), (**c**) & (**e**) in turn but for the AFM case ($K_0 = -1/8$). The renormalized parameters for the FM and AFM cases have different winding numbers (determined by how many periods ($2\pi$) of time the parameter curves are spanned over).

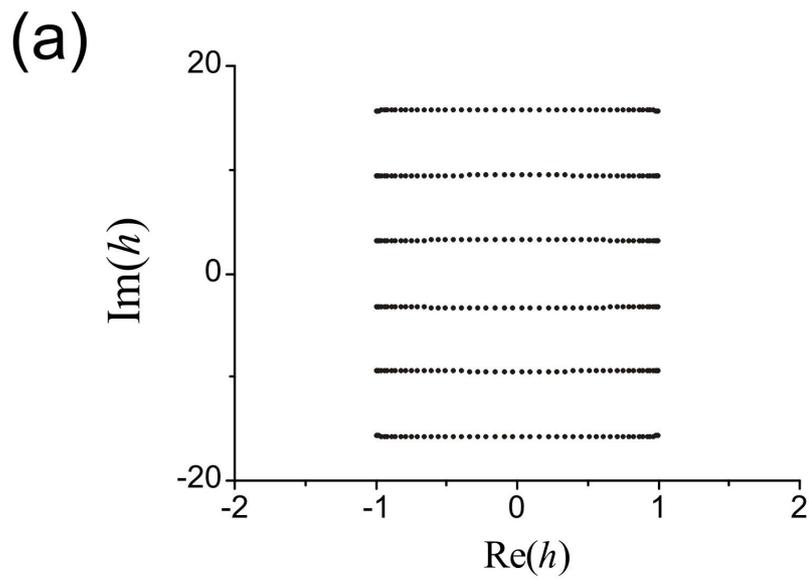

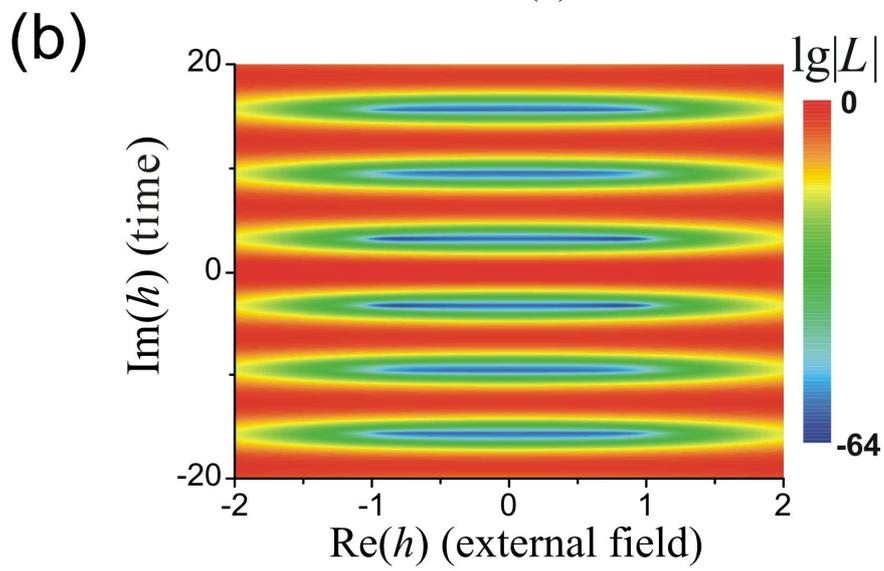

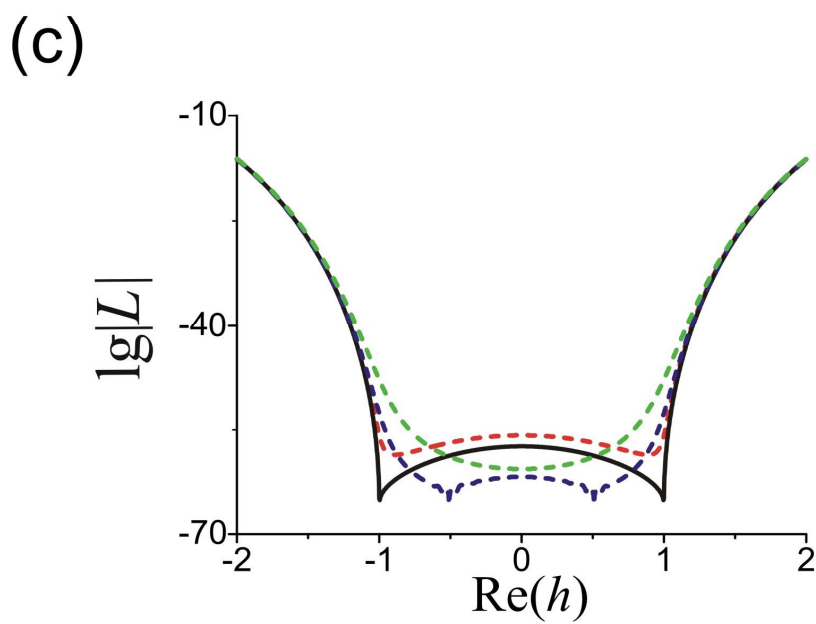

**Figure 6| Finite-temperature phase transitions of a transverse field Ising spin chain revealed by time-domain observation.** The one-dimensional model contains 100 spins. The temperature is such that $\beta = 0.5$. **a.** Lee-Yang zeros in the complex plane of the external field. **b**. Contour plot of the time-domain measurement, $\lg|L|$, as a function of the external field Re($h$) and the time Im($h$). **c.** $\lg|L|$ plotted as functions of the external field Re($h$) for different times Im($h$) = 3.10 (red-dashed line), 3.14 (black-solid line), 3.26 (blue-dashed line) and 3.35 (green-dashed line). The phase transitions at Re($h$) = ±1 are observed.

# SUPPLEMENTARY INFORMATION
## for
## Phase transitions in the complex plane of physical parameters

## I. Full Methods and Supplementary Equations

### A. Exact solution and Lee-Yang zeros of the 1D Ising model

The Hamiltonian of the 1D Ising model at finite external magnetic field is

$$H(h) = -\sum_{j=1}^{N}\left(J\sigma_j\sigma_{j+1} + h\sigma_j\right). \quad (S1)$$

The partition function can be exactly obtained by the transfer matrix method as

$$Z = e^{N\beta J}\left[\left(\cosh(\beta h) + \sqrt{\sinh^2(\beta h) + e^{-4\beta J}}\right)^N + \left(\cosh(\beta h) - \sqrt{\sinh^2(\beta h) + e^{-4\beta J}}\right)^N\right]. \quad (S2)$$

With $z \equiv \exp(-2\beta h)$ and $x \equiv \exp(-2\beta J)$, the partition function is written as

$$Z = e^{N(\beta J + \beta h)}\left(\frac{1+z}{2} + \sqrt{\left(\frac{1-z}{2}\right)^2 + x^2 z}\right)^N \left[1 + \left(\frac{\frac{1+z}{2} - \sqrt{\left(\frac{1-z}{2}\right)^2 + x^2 z}}{\frac{1+z}{2} + \sqrt{\left(\frac{1-z}{2}\right)^2 + x^2 z}}\right)^N\right]. \quad (S3)$$

Therefore the Lee-Yang zeros are given by

$$\frac{\frac{1+z}{2} - \sqrt{\left(\frac{1-z}{2}\right)^2 + x^2 z}}{\frac{1+z}{2} + \sqrt{\left(\frac{1-z}{2}\right)^2 + x^2 z}} = e^{ik_n}, \quad (S4)$$

where $k_n = \pi(2n-1)/N, n = 1, 2, \cdots, N$. After solving the above algebraic equation of $z$, the Lee-Yang zeros for ferromagnetic coupling ($J > 0$) are

$$\begin{aligned} z_n &= \cos\theta_n + i\sin\theta_n \\ \cos\theta_n &= -x^2 + (1-x^2)\cos k_n, \\ \sin\theta_n &= \pm\sqrt{(1-x^2)[\sin^2 k_n + x^2(1+\cos k_n)^2]}. \end{aligned} \quad (S5)$$

For anti-ferromagnetic phase, $J < 0$,

$$z_n = -x^2 + (1-x^2)\cos k_n \pm \sqrt{(x^2-1)[\sin^2 k_n + x^2(1+\cos k_n)^2]}. \quad (S6)$$

For non-interacting case, $J = 0$,

$$z_n = -1. \quad (S7)$$

## B. 2D Ising model in a finite field

The Hamiltonian of the 2D square lattice Ising model with a finite magnetic field is

$$H = -\sum_{\langle i,j \rangle} J\sigma_i\sigma_j - h\sum_j \sigma_j \quad (S8)$$

The 2D square lattice Ising model was solved exactly by Onsager in the zero field. For non-zero field, there is no exact solution, but one can map the 2D classical spin model to a 1D quantum spin model by the transfer matrix method:

$$\begin{aligned} Z(\beta, h) &= \mathrm{Tr}\left[ e^{\beta J \sum_{i=1,j=1}^{M,N} \sigma_{i,j}\sigma_{i,j+1} + \beta J \sum_{i=1,j=1}^{M,N} \sigma_{i,j}\sigma_{i+1,j} + \beta h \sum_{i=1,j=1}^{M,N} \sigma_{i,j}} \right] \\ &= \sum_{\{\mu_j\},\{\mu'_j\}} \left[ C_{\mu_1\mu_1'} D_{\mu_1'\mu_2} \right]\left[ C_{\mu_2\mu_2'} D_{\mu_2'\mu_3} \right]\cdots\left[ C_{\mu_M\mu_M'} D_{\mu_M'\mu_1} \right] \\ &= \mathrm{Tr}[(CD)^M], \end{aligned} \quad (S9)$$

where

$$C_{\mu\mu'} \equiv e^{\beta J \sum_{j=1}^{N} \mu(j)\mu'(j+1) + \beta h \sum_{j=1}^{N} \mu'(j)} \delta_{\mu\mu'} \text{ and } D_{\mu\mu'} \equiv e^{\beta J \sum_{j=1}^{N} \mu(j)\mu'(j)} = e^{\beta J s_1 s_1'} e^{\beta J s_2 s_2'} \cdots e^{\beta J s_N s_N'}.$$

Therefore, evaluation of the partition function of the 2D Ising model amounts to solution of a 1D quantum spin chain:

$$Z(\beta,h)=\text{Tr}[(CD)^M],$$
$$C = \exp\left[\beta J \sum_{j=1}^{N} \sigma_j^z \sigma_{j+1}^z + \beta h \sum_{j=1}^{N} \sigma_j^z\right],$$
$$D = (e^{\beta J} + e^{-\beta J}\sigma_1^x)(e^{\beta J} + e^{-\beta J}\sigma_2^x)\cdots(e^{\beta J} + e^{-\beta J}\sigma_N^x)$$
$$= [2\sinh(2K)]^{N/2} \exp\left[K^* \sum_{j=1}^{N} \sigma_j^x\right].$$
(S10)

## C. Exact RG flow equations for 1D Ising model

Denoting the dimensionless parameters $K_0 = \beta J$ and $h_0 = \beta h$ for the Hamiltonian in equation (S1), one can evaluate the partition function as [12]

$$\begin{aligned}
Z(\tilde{J},\tilde{h};N) &= \text{Tr}\left[\exp\left(\sum_{j=1}^{N}(K_0\sigma_j\sigma_{j+1} + h_0\sigma_j)\right)\right] \\
&= \text{Tr}\left[\prod_{j\text{ even}}^{N/2}\left(\sum_{\sigma_{j+1}=\pm 1}\exp\left(K_0(\sigma_j\sigma_{j+1}+\sigma_{j+1}\sigma_{j+2})+\frac{h_0}{2}(\sigma_j+2\sigma_{j+1}+\sigma_{j+2})\right)\right)\right] \\
&= \text{Tr}\left[\prod_{j\text{ even}}^{N/2}\exp\left(K_1\sigma_j\sigma_{j+2}+\frac{h_1}{2}(\sigma_j+\sigma_{j+2})+G_1\right)\right] \\
&= e^{NG_1/2}Z(K_1,h_1;N/2),
\end{aligned}$$
(S11)

where we have assumed periodic boundary condition and $N$ to be even for simplicity. This gives the RG flow equations

$$\begin{aligned}
K_1 &= \frac{1}{4}\ln\left(\frac{\cosh(2K_0+h_0)\cosh(2K_0-h_0)}{\cosh^2(h_0)}\right) \\
h_1 &= h_0 + \frac{1}{2}\ln\left(\frac{\cosh(2K_0+h_0)}{\cosh(2K_0-h_0)}\right);\ \& \\
G_1 &= \frac{1}{4}\ln\left(16\cosh^2(h_0)\cosh(2K_0+h_0)\cosh(2K_0-h_0)\right).
\end{aligned}$$
(S12)

## D. Approximate RG flow equations for 2D Ising model

Denoting the dimensionless parameters $K_0 = \beta J$ and $h_0 = \beta h$ for the Hamiltonian in equation (S8). The partition function of 2D Ising model can be evaluated

$$Z(\beta,h) = \mathrm{Tr}\left[\exp\left(K_0\sum_{i,j=1}^{N}\sigma_{i,j}\sigma_{i,j+1} + K_0\sum_{i,j=1}^{N}\sigma_{i,j}\sigma_{i+1,j} + h_0\sum_{j=1}^{N}\sigma_{i,j}\right)\right]$$
$$= e^{-NG_1/2}\mathrm{Tr}\left[\exp\left(\begin{array}{c}h_1\sum_p\sigma_p + K_1\sum_{nn}\sigma_p\sigma_q \\ +K_2\sum_{nnn}\sigma_p\sigma_q + K_3\sum_{pqr}\sigma_p\sigma_q\sigma_r + K_4\sum_{pqrs}\sigma_p\sigma_q\sigma_r\sigma_s\end{array}\right)\right],$$
(S13)

where nn denotes nearest neighbor spins and nnn denotes next nearest neighbor spins in the renormalized lattice. The renormalized parameters are

$$K_1 = \frac{1}{8}\ln\left(\frac{\cosh(4K_0+h_0)\cosh(4K_0-h_0)}{\cosh^2(h_0)}\right),$$
$$K_2 = \frac{1}{16}\ln\left(\frac{\cosh(4K_0+h_0)\cosh(4K_0-h_0)}{\cosh^2(h_0)}\right),$$
$$K_3 = \frac{1}{16}\ln\left(\frac{\cosh(4K_0+h_0)\cosh^2(2K_0-h_0)}{\cosh(4K_0-h_0)\cosh^2(2K_0+h_0)}\right), \quad (S14)$$
$$K_4 = \frac{1}{16}\ln\left(\frac{\cosh(4K_0+h_0)\cosh(4K_0-h_0)\cosh^6(h_0)}{\cosh^4(2K_0+h_0)\cosh^4(2K_0-h_0)}\right),$$
$$h_1 = h_0 + \frac{1}{4}\ln\left(\frac{\cosh(4K_0+h_0)\cosh^2(2K_0+h_0)}{\cosh(4K_0-h_0)\cosh^2(2K_0-h_0)}\right).$$

The RG equations are not closed as new parameters $K_2, K_3, K_4$ emerge. It is necessary to make some approximation to make the RG equations close. A truncation scheme [13] is obtained by correcting the theory in an approximate way for the presence of the term $K_2, K_3, K_4$. Since both $K_2$ and $K_4$ are positive, the emergent interaction associated with them have the effect of increasing the alignment of spins. Hence a reasonable approximation is to drop $K_2, K_3, K_4$ but simultaneously increase $K_1$ to a new value so that the new alignment tendency remains the same. Thus we have the approximate RG equations for 2D Ising model,

$$K_1' \approx \frac{3}{16}\ln\left(\frac{\cosh(4K_0+h_0)\cosh(4K_0-h_0)}{\cosh^2(h_0)}\right),$$
$$h_1 = h_0 - \frac{i}{4}\ln\left(\frac{\cosh(4K_0+h_0)\cosh^2(2K_0+h_0)}{\cosh(4K_0-h_0)\cosh^2(2K_0-h_0)}\right).$$
(S15)

### E. Exact solution and Lee-Yang zeros of the one-dimensional transverse field Ising model.

The Hamiltonian of the one-dimensional transverse-field Ising model is

$$H = \lambda_1 H_1 + \lambda_2 H_2,$$
$$H_1 = \sum_{j=1}^{N} \sigma_j^x \sigma_{j+1}^x, \quad H_2 = \sum_{j=1}^{N} \sigma_j^z, \tag{S16}$$

which can be exactly solved [14]. By applying Jordan-Wigner transformation, Fourier transformation and Bogoliubov transformation, the Hamiltonian is transformed to be a free-fermion one as

$$H = -\lambda_1 \sum_k \varepsilon(k)\left(b_k^\dagger b_k - 1/2\right), \tag{S17}$$

with energy dispersion

$$\varepsilon(k) = 2\sqrt{\left(\cos k - \frac{\lambda_2}{\lambda_1}\right)^2 + \sin^2 k}. \tag{S18}$$

Therefore, at temperature $T$, the positions of the zeros of the partition function $Z = \text{Tr}\left(e^{-\beta H}\right)$ are decided by

$$\exp\left(-\beta \lambda_1 \varepsilon(k)/2\right) + \exp\left(\beta \lambda_1 \varepsilon(k)/2\right) = 0, \tag{S19}$$

i.e.

$$2\beta\lambda_1 \sqrt{\left(\cos k - \frac{\lambda_2}{\lambda_1}\right)^2 + \sin^2 k} = i(2n+1)\pi, \tag{S20}$$

with $n$ being an integer.

### F. Physical realizations for the time-domain measurement.

The time-domain measurement may be implemented by coupling the system to a probe spin through the probe-system coupling $H_{SB} = |\uparrow\rangle\langle\uparrow| \otimes H_\uparrow(t) + |\downarrow\rangle\langle\downarrow| \otimes H_\downarrow(t)$ ($S_z \equiv |\uparrow\rangle\langle\uparrow| - |\downarrow\rangle\langle\downarrow|$) and measuring the probe spin coherence. If we initialize the probe spin in a superposition state $|\uparrow\rangle+|\downarrow\rangle$ and the system in a thermal equilibrium state described by the density matrix, $\rho=\exp(-\beta H)/\text{Tr}[\exp(-\beta H)]$, the coherence function of the quantum probe is,

$$\langle S_x\rangle+i\langle S_y\rangle=\frac{\text{Tr}[e^{-i(H+H_\downarrow)t}e^{-\beta H}e^{i(H+H_\uparrow)t}]}{\text{Tr}[e^{-\beta H}]}, \tag{S21}$$

(1). If $[H_I,H]=0$, the time-domain measurement can be reduced to $L(t)=Z^{-1}\text{Tr}\left[\exp(-\beta H)\exp(-iH_I t)\right]$, which can be implemented by probe spin decoherence with a time-independent probe-bath coupling, i.e. $H_\uparrow=-H_\downarrow=H_I/2$.

(2). If $[H_I,H]\neq 0, [[H_I,H],H]=[[H_I,H],H_I]=0$, the time-domain measurement can be written as

$$\begin{aligned}L(t)&=\frac{\text{Tr}\left[\exp(-\beta H-iH_I t)\right]}{\text{Tr}\left[\exp(-\beta H)\right]}\\&=\frac{\text{Tr}\left[\exp(-\beta H/2-iH_I t/2)\exp(-\beta H/2-iH_I t/2)\right]}{\text{Tr}\left[\exp(-\beta H)\right]}\\&=\frac{\text{Tr}\left[e^{i\beta t[H,H_I]/4}e^{-iH_I t/2}e^{-\beta H/2}e^{-\beta H/2}e^{-iH_I t/2}e^{-i\beta t[H,H_I]/4}\right]}{\text{Tr}\left[e^{-\beta H}\right]}\\&=\frac{\text{Tr}\left[\exp(-iH_I t/2)\exp(-\beta H)\exp(-iH_I t/2)\right]}{\text{Tr}\left[\exp(-\beta H)\right]}.\end{aligned} \tag{S22}$$

Comparing with equation (S21), we can implement the time-domain measurement by probe spin decoherence with a time-independent probe-bath coupling $H_\downarrow=H_I/2-H\ \&\ H_\uparrow=-H_I/2-H$.

(3). If $[H_I,H]\neq 0, [H,H_I]=isH_I$, with $s$ a real number, the time-domain measurement can be written as

$$L(t) = \frac{\text{Tr}\left[\exp(-\beta H - iH_I t)\right]}{\text{Tr}\left[\exp(-\beta H)\right]}$$

$$= \frac{\text{Tr}\left[\exp(-\beta H/2 - iH_I t/2)\exp(-\beta H/2 - iH_I t/2)\right]}{\text{Tr}\left[\exp(-\beta H)\right]}$$

$$= \frac{\text{Tr}\left[e^{-(e^{is}-1)H_I t/2s} e^{-\beta H/2} e^{-\beta H/2} e^{-(1-e^{-is})H_I t/2s}\right]}{\text{Tr}\left[e^{-\beta H}\right]} \qquad \text{(S23)}$$

$$= \frac{\text{Tr}\left[\exp(-it\sin(s)H_I/2s)\exp(-\beta H)\exp(-it\sin(s)H_I/2s)\right]}{\text{Tr}\left[\exp(-\beta H)\right]}.$$

Comparing with equation (S21), we can implement the time-domain measurement by probe spin decoherence with a time-independent probe-bath coupling $H_\downarrow = \sin(s)H_I/2s - H$ & $H_\uparrow = -\sin(s)H_I/2s - H$.

(4). In the more general cases where $[H_I, H] \neq 0$, the time-domain measurement in equation(1) of the main text can be written as $L(t) = \text{Tr}\left[e^{-\beta H'} e^{-itH_I'}\right] / \text{Tr}\left[e^{-\beta H}\right]$, where

$$e^{-2\beta H'} \equiv T \exp\left(-\frac{\beta}{2}\int_{-1}^{1} \tilde{H}(u)\, du\right) \bar{T} \exp\left(-\frac{\beta}{2}\int_{-1}^{1} \tilde{H}(u)\, du\right),$$

$\exp(-itH_I') \equiv \exp(\beta H')\exp(-\beta H - itH_I)$, $\tilde{H}(u) \equiv \exp(iutH_I) H \exp(-iutH_I)$ and $T$ and $\bar{T}$ are the time-ordering and anti-ordering operators, respectively. If one initializes the bath in a canonical state $\rho = \exp(-\beta H')/\text{Tr}[\exp(-\beta H')]$, the time-domain measurement can be implemented by probe spin decoherence with a probe-bath coupling $H_\downarrow = H_I'/2 - H'$ & $H_\uparrow = -H_I'/2 - H'$ up to a normalization factor

$$L(t) = \frac{\text{Tr}\left[\exp(-\beta H - iH_I t)\right]}{\text{Tr}\left[\exp(-\beta H)\right]}$$

$$= \frac{\text{Tr}\left[e^{-\beta H'} e^{-itH_I'}\right]}{\text{Tr}\left[e^{-\beta H'}\right]} \times \frac{\text{Tr}\left[e^{-\beta H'}\right]}{\text{Tr}\left[e^{-\beta H}\right]}. \qquad \text{(S24)}$$